# Addressing learning difficulties in Newton's 1st and 3rd Laws through problem based inquiry using Easy Java Simulation

Khoon Song Aloysius GOH[1], Loo Kang WEE[2], Kim Wah YIP[1], Ping Yong Jeffrey TOH[1] and Sze Yee LYE[2]

[1]Ministry of Education, Anderson Junior College, Singapore
[2]Ministry of Education, Education Technology Division, Singapore

goh_khoon_song@moe.edu.sg, wee_loo_kang@moe.gov.sg, yip_kim_wah@moe.edu.sg, toh_ping_yong_jeffrey@moe.edu.sg, lye_sze_yee@moe.gov.sg

We develop an Easy Java Simulation (EJS) model for students to visualize Newton's 1st and 3rd laws, using frictionless constant motion equation and a spring collision equation during impact. Using Physics by Inquiry instructional (PbI) strategy, the simulation and its problem based inquiry worksheet aim to enhance learning of these two Newtonian concepts. We report results from Experimental (N=62 students) and Control (N=67) Groups in 11 multiple-choice questions pre- and post-tests, conducted by three teachers in the school. Results suggest, at 95% confidence level, significant improvement for concept of Newton's 1st Law while not so for Newton's 3rd Law. A Focus Group Discussion revealed students confirming the usefulness of the EJS model in visualizing the 1st Law while not so much for the 3rd Law. We speculate the design ideas for constant velocity motion in the computer model coupled with the PbI worksheet did allow for 'making sense' and experiencing of the 1st Law, where traditional pen-paper representations could not. We have improved the features for the action-reaction contact forces visualization associated with the 3rd Law and we hope other teachers will find the simulation useful for their classes and further customize them to benefit all mankind, becoming citizens for the world.

PACS: 45.10.-b  45.20.df  45.20.dh  01.40.-d  01.50.H-

## I. INTRODUCTION

Physics by Inquiry (PbI) as an instructional strategy in a local school has gained popularity, but results from the localized version of the Force Concept Inventory (FCI) suggests students still harbor commonsense beliefs about motion with forces (Halloun & Hestenes, 1985), inconsistent with Newton's 1st and 3rd laws. This is probably due to a combination of many factors, one of the main causes is the difficulty to "make sense" (Wee, 2012a) of the phenomena, without learning by first person experiencing (Oblinger, 2004; Wee, 2012b) and contextualizing in "real-life referents" (Dede, Salzman, Loftin, & Sprague, 1999), hence leading to what is commonly referred to as the abstract nature (Chabay & Sherwood, 2006) of learning physics.

We argue that computer simulation could be an appropriate substitute for active learning referents, provided simulations are carefully developed (Weiman & Perkins, 2005), used in appropriate context (Finkelstein et al., 2005), aided with challenging inquiry activities (Adams, Paulson, & Wieman, 2008) and facilitated by teachers who believe (Hsu, Wu, & Hwang, 2007) in the effectiveness of the tool.

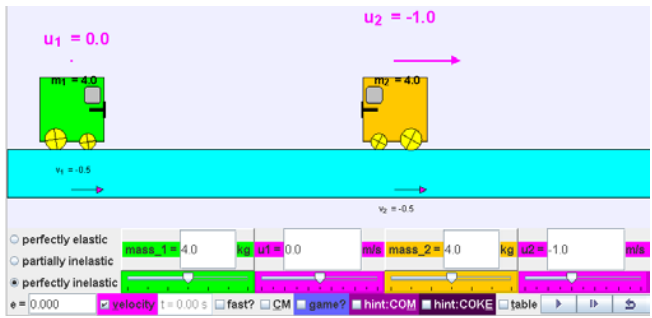

Figure 1. EJS applet view of the virtual laboratory simulation learning environment showing a world view, and a bottom control panel for student-directed inquiry activities where students are able to make sense of Newton's 1st and 3rd Laws.

Building on open source codes shared by the Open Source Physics (OSP) community like, Francisco's example of "Collision in one dimension" (Esquembre, 2009), Andrew's (Duffy, 2010) One Dimensional Collision Model for game design, and Fu-Kwun's many other examples on NTNUJAVA Virtual Physics Laboratory, we further customize an Easy Java Simulation (EJS) (Wee & Esquembre, 2008) computer model into a virtual laboratory as shown in Figure 1 (Wee, Esquembre, & Lye, 2012), that we hope many teachers will find useful and can act more intelligibly (Juuti & Lavonen, 2006) in their own classes.

## II. PHYSICS MODEL

In this simulation, the two-body collision carts model is simulated by constant velocities motion as equations (1) and (2), assuming the $x$ position of the centre of carts 1 and 2 are $x_1$ and $x_2$ respectively and their instantaneous velocities $v_1$ and $v_2$ respectively (see Figure 2).

$$\frac{dx_1}{dt} = v_1 \qquad (1)$$

$$\frac{dx_2}{dt} = v_2 \qquad (2)$$

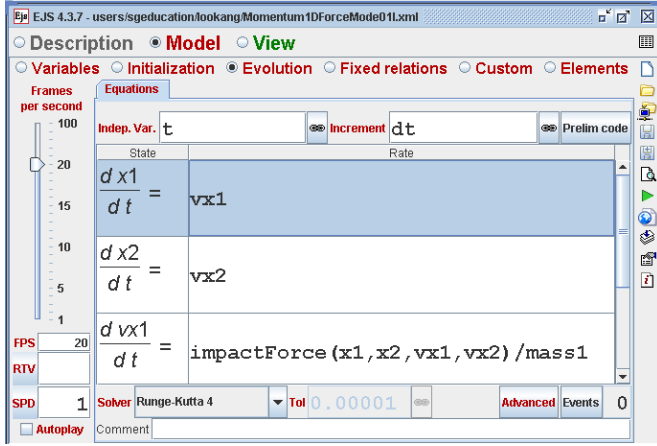

Figure 2. EJS authoring tool view at the 'Evolution Page' showing equations (1) and (2) as ordinary differential equations (ODE) with time $t$ as the independent variable and $dt$ as the increment.

Notice how easily these equations simulate carts that continue in uniform $x$ direction motion without any loss of energy as described by Newton's $1^{st}$ Law.

The contact impact force is modeled by equation (3) as adapted from Brach (2003, p. 3) where $k$ is a linear spring constant, $es$ is the coefficient of restitution, $m_1$ and $m_2$ are masses of carts 1 and 2 respectively.

$$F_{impact} = -2\sqrt{\frac{\log(es)^2}{\pi^2 + \log(es)^2}}\sqrt{k(\frac{m_1 m_2}{m_1 + m_2})}(v_1 - v_2) + k(x_1 - x_2)$$

(3)

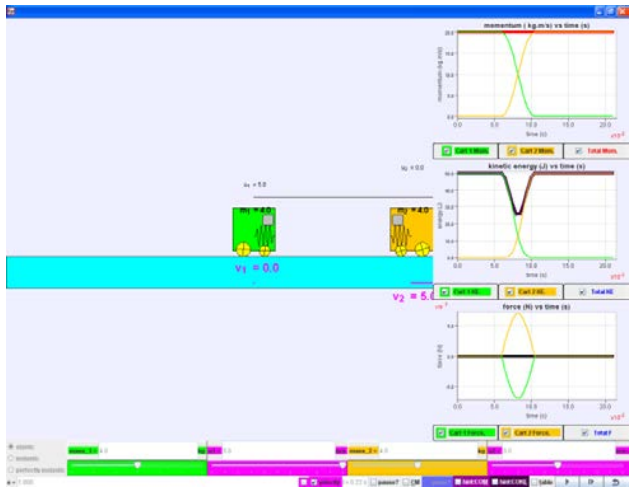

Figure 3. Collision carts (realistic) model (Wee, Esquembre, et al., 2012) derived from Francisco's original work (Esquembre, 2009) with three scientific graphs showing realistic spring modelled during collisions.

This Physics model (see Figure 3), when implemented in a simulation, allows experiencing and 'messing about' productively (Finkelstein, et al., 2005, pp. 010103-010107); and serving as a powerful referent tool (Dede, et al., 1999) for learning.

## III. METHOD

This study investigates whether students from the experimental group who have undergone the PbI problem based inquiry lesson using a finer customized EJS computer model will have a better learning experience than their peers in the control traditional-teaching group. Our team of three teachers each selects two of their classes to participate in this research study. The classes are assigned with the intention of creating equivalent groups of similar class size and similar mean subject grade of 2.00 equivalent of 'B' grade (Table I) in their Ordinary Level Physics. The same teacher participating in both groups serves to reduce the instructor effect.

Table I. Class sizes of Experimental and Control Groups of the instructors. Mean Subject Grades of Experimental and Control Groups are similar.

| Instructor | Experimental Group (EG) | Control Group (CG) |
|---|---|---|
| YKW | 23 | 24 |
| AG | 22 | 21 |
| JT | 17 | 22 |
| Total | 62 | 67 |
| Mean Subject Grade (MSG) | 2.00 = 'B' | 2.00 = 'B' |

Prior to attending lessons on the topic of "Dynamics", the entire cohort of about 400 Physics students in this school took a pre-test based on a selection of 15 questions from the Force Concept Inventory (FCI), which focuses particularly on Newton's three Laws. Out of the 15 questions, our test data was collected from seven multiple-choice questions (MCQ) on Newton's First Law (N1stL) and four MCQ on Newton's Third Law (N3rdL). After the topic was completed after 2 to 3 weeks, the students took a post test, identical to the pre-test. Furthermore, two questions about N1stL and N3rdL were specially crafted for their Mid-Year Common Test to assess the longer term transfer of their conceptual change.

Focus Group Discussion (FGD) with 9 students, three students from each experimental class was conducted where they were to reflect on their learning experience so that the lesson package of worksheet and computer model can be further improved.

## IV. RESULTS AND DISCUSSIONS

Using Z-test as in equation (4), for Newton's $1^{st}$ Law, the Z-value is 2.23 where $|Z| > 1.96$ (see Table II). There is sufficient evidence at 5% significance level to reject the null hypothesis that the Experimental Group did not do better than the Control Group for Newton's $1^{st}$ Law.

$$Z = \frac{(\overline{X_1} - \overline{X_2}) - (\mu_1 - \mu_2)}{\sqrt{\frac{\sigma_1^2}{N_1} + \frac{\sigma_2^2}{N_2}}}$$

(4)

Table II. Pre-and post-tests scores categorized according to question type; Newton's $1^{st}$ Law (7 marks) and $3^{rd}$ Law (4 marks) for the Experimental ($N_1$=62) and Control ($N_2$=67) Groups.

| Question Type | Group | Pre-test score $\mu$ | Post-test score X | Difference (X-$\mu$)±σ | Probability value |
|---|---|---|---|---|---|
| N1stL (7 marks max) | $N_1=$ 62 | $3.968 \pm 1.736$ | $4.774 \pm 1.562$ | $0.806 \pm 2.055$ | P(|Z| < 2.23) |
| | $N_2=$ 67 | $4.612 \pm 1.477$ | $4.716 \pm 1.485$ | $0.104 \pm 1.447$ | =0.974 |
| N3rdL (4 marks max) | $N_1=$ 62 | $2.065 \pm 1.143$ | $2.823 \pm 1.138$ | $0.758 \pm 1.224$ | P(|Z| < 0.76) |
| | $N_2=$ 67 | $2.060 \pm 1.071$ | $2.970 \pm 0.953$ | $0.910 \pm 1.041$ | =0.552 |

However, for Newton's $3^{rd}$ Law, the Z-value is -0.76 where $|Z| < 1.96$. There is insufficient evidence at 5% significance level to reject the null hypothesis that Experimental Group did not do better than the Control Group for Newton's $3^{rd}$ Law.

In addition, Experimental Group continued to perform better in Newton's $1^{st}$ Law in the Mid-year common test. However, the Experimental Group did not perform better in Newton's $3^{rd}$ Law (Table III).

Table III. Mid-year common test scores categorized according to question type; Newton's $1^{st}$ Law (2 marks) and $3^{rd}$ Law (2 marks) for the Experimental ($N_1=62$) and Control ($N_2=67$) Groups.

| Question Type | Group | Mid-year test score $Y$ |
|---|---|---|
| N1stL (2 marks max) | $N_1=62$ | $1.032 \pm 0.829$ |
| | $N_2=67$ | $0.836 \pm 0.853$ |
| N3rdL (2 marks max) | $N_1=62$ | $0.032 \pm 0.254$ |
| | $N_2=67$ | $0.164 \pm 0.559$ |

This gap in the learning of the $3^{rd}$ Law has allowed for the implementation of new design idea-features such as clearer and slow motion visualization of contact forces during collision and larger mass having proportional larger area representation, which we will test out in future research lessons (see Figure 4).

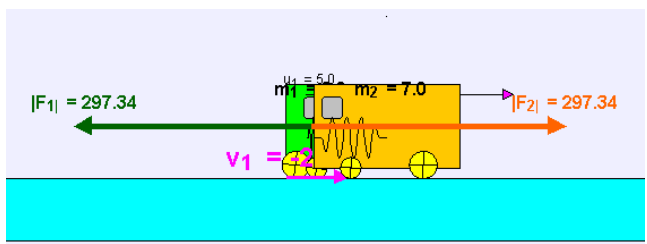

Figure 4. Collision carts (realistic) model (Wee, Esquembre, et al., 2012) with design feature-ideas to bring out the concepts of Newton's $3^{rd}$ Law where the contact forces are clearly represented as equal, opposite and acting on different bodies even for masses that are different.

We include excerpts of the Focus Group Discussion and informal interviews with the students to give some themes and insights into the conditions and processes during the problem based inquiry lessons. Words in brackets [] are added to improve the readability of the qualitative interviews.

### 1) Computer model allows visualization

"This is more for visual learners. They can see how they actually collide, in which direction and what will be the results."

"The computer model will help you to see the [representation of] forces acting at any one instant, unlike the real collision carts demonstration set [which are invisible]."

### 2) Need for balance between student direct inquiry and teacher direct instruction

"Overall I feel that this project is useful because it enforces self-exploration of the interactions between colliding objects and this is especially useful for those with inquisitive minds as they are able to configure the velocity as well as the type of collision."

"With teacher demonstrating the use of computer models, logistically more efficient, but learning wise may not be better [because student need to direct the inquiry to deepen their understanding]"

### 3) Need for real equipment to relate to the real world

"Actually the programme [computer model] does help in some ways, but ……. we [still] can't really relate this to real life situations."

"I think it's quite closely related to our syllabus and can use this live experiment to foster a deeper understanding of how the force works (in collision carts)."

Readers could explore use of Tracker (Brown, 2012; Wee, Chew, Goh, Tan, & Lee, 2012; Wee & Lee, 2011) to allow students to inquire into videos of real collisions for a stronger connection of scientific concepts to real life applications.

## V. CONCLUSION

The theoretical physics model of a two-body realistic collision system in one dimension is discussed and implemented in EJS and the equations (1) to (3) should be applicable to any modeling tool such as VPython (Scherer, Dubois, & Sherwood, 2000) or Modellus (Teodoro, 2004).

Our research data using Z-test suggests that at a 95% confidence level, students who underwent treatment of PbI worksheet with our customized EJS computer model (N=62) performed better in Newton's $1^{st}$ Law than their peers who otherwise underwent the traditional instructions (N=67). Focus Group Discussion with students and discussions with teachers suggest the computer model design and its pedagogical use as a tool did allow students to 'make sense' and experience the $1^{st}$ Law.

We hope this paper adds to the body of knowledge when computer models are used for interactive engagement (Hake, 1998) by the students. We also hope that computer models can provide experiential learning (Wee, 2012b) and sensing; making visualization better which is not possible through traditional paper media.

We have since improved on the computer model for better visualization of the $3^{rd}$ Law with design ideas as shown (see Figure 4).

We hope teachers will find the worksheet and computer model (https://www.dropbox.com/s/gf1vc7qqy7l47y8/CollsionCarts AJCworksheets.zip and

https://www.dropbox.com/s/8sgjazk5dohj5sk/ejs_Momentum 1DForceModel01.jar) useful and can act more intelligibly (Juuti & Lavonen, 2006) in their own classes.

## VI. ACKNOWLEDGEMENT

We wish to acknowledge the passionate contributions of Francisco Esquembre, Fu-Kwun Hwang, Wolfgang Christian, Andrew Duffy, Todd Timberlake and Juan Aguirregabiria for their ideas and insights in the co-creation of interactive simulation and curriculum materials.

This research is made possible thanks to the eduLab project NRF2011-EDU001-EL001 Java Simulation Design for Teaching and Learning, (MOE, 2012b) awarded by the National Research Foundation (NRF), Singapore in collaboration with National Institute of Education (NIE), Singapore and the Ministry of Education (MOE), Singapore.

We also thank MOE for the recognition of our research on computer model lessons as a significant innovation in 2012 MOE Innergy (HQ) GOLD Awards (MOE, 2012a) by Educational Technology Division and Academy of Singapore Teachers.

Any opinions, findings, conclusions or recommendations expressed in this paper, are those of the authors and do not necessarily reflect the views of the MOE, NIE or NRF.

AUTHOR

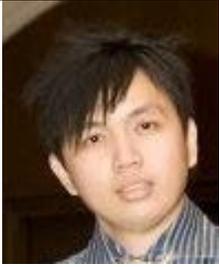

Khoon Song Aloysius GOH is a physics teacher in Anderson Junior College. He holds a Bachelor's degree in mechanical engineering from National University of Singapore and a Master's degree in business administration (MBA) from University of Strathclyde. His academic and professional interests include the appropriate use of ICT to enhance learning and feasibility of organization management theories in Singapore's public school system.

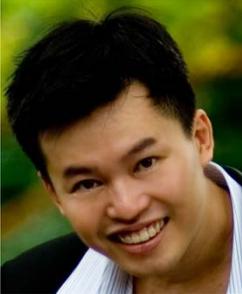

Loo Kang Lawrence WEE is currently an educational technology specialist at the Ministry of Education, Singapore. He was a junior college physics lecturer and his research interest is in Open Source Physics tools like Easy Java Simulation for designing computer models and use of Tracker.

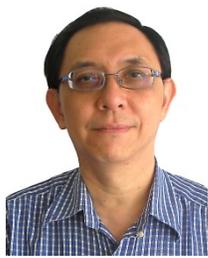

Kim Wah YIP has been teaching Physics at Anderson Junior College since his first posting in 1985. He holds a Master of Science degree in Physics by research on "Computer Simulation of High Temperature Semiconductors" from the National University of Singapore.

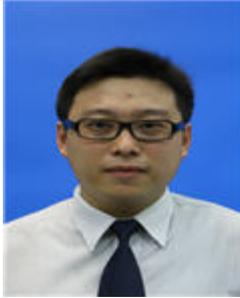

Jeffrey TOH has been teaching Physics at Anderson Junior College since 2010. He holds a joint Master of Science degree in Atomic Scale Modelling of Physical, Chemical and Biomolecular Systems from the Ecole Normale Superieure de Lyon and Universiteit van Amsterdam. His research interests include innovative pedagogies to help students learn H1 and H2 Physics better.

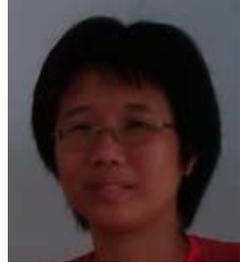

Sze Yee Lye is currently an educational technology officer in Ministry of Education, Singapore. She is a trained Physics Teacher and had taught both Physics and science in secondary and primary schools. She is now working on modifying the Open Source Physics Simulations for physics-related topics in primary school.